\documentstyle[12pt]{article}
\font\big = cmbx12 scaled \magstep3

\font\sm = cmr10
\font\bsm = cmbx10
\font\tbf = cmbx12 scaled \magstep2
\begin{document}
%%%%%%%%%%%%%%%%%%%%%%%%%%%%%%%%%%%%%%%%%%%%%%%%%%%%%%%%%%%%%%%%%%%%
%%%%%%%%%%%%%% TITLEPAGE AND ABSTRACT %%%%%%%%%%%%%%%%%%%%%%%%%%%%%%
%%%%%%%%%%%%%%%%%%%%%%%%%%%%%%%%%%%%%%%%%%%%%%%%%%%%%%%%%%%%%%%%%%%%
\centerline{\big Ternary generalizations of Grassmann}
\vskip .2cm
\centerline{\big algebra} 
\vskip .8cm
%%%%%%%%%%%%%%%%%%%%%%%%%%%%%%%%%%%%%%%%%%%%%%%%%%%%%%%%%%%%%%%%%%%%
\centerline{\bf Viktor Abramov}

\vskip .5cm
\noindent
{\sm Puhta Matemaatika Instituut, Tartu \"Ulikool (Institute of Pure
Mathematics, University of Tartu), Vanemuise 46, Tartu, Eesti (Estonia).}

\vskip .3cm
%%%%%%%%%%%%%%%%%%%%%%%%%%%%%%%%%%%%%%%%%%%%%%%%%%%%%%%%%%%%%%%%%%%%
\noindent
{\bsm Abstract.} {\sm We propose the ternary generalization of the classical
anti-commutativity and study the algebras whose generators are ternary
anti-commutative. The integral over an algebra with an arbitrary
number of generators $N$ is defined and the formula of a change of
variables is proved.
In analogy with the fermion integral we define an analogue of the
Pfaffian for a cubic matrix by means of Gaussian type integral and
calculate its explicit form in the case of $N=3$. }
%%%%%%%%%%%%%%%%%%%%%%%%%%%%%%%%%%%%%%%%%%%%%%%%%%%%%%%%%%%%%%%%%%%%%%%%%%%%

\vskip .3cm
\noindent
{\bsm Keywords:} {\sm Ternary anti-commutativity, Grassmann algebra,
Pfaffian.} 
%%%%%%%%%%%%%%%%%%%%%%%%%%%%%%%%%%%%%%%%%%%%%%%%%%%%%%%%%%%%%%%%%%%%%%%%%%%

\vskip .8cm
\centerline{\tbf 1 Introduction}
\vskip .3cm
\indent
If ${\cal A}$ is an algebra with composition law $(a,b)\to a\cdot b$
then its composition law is said to be anti-commutative if $a^2=0,
\forall a\in {\cal A}$. The most known examples of an algebra with
anti-commutative multiplication are provided by Lie algebras. The first
natural generalization of anti-commutative multiplication is to increase
the number of arguments, i.e. to consider the algebras whose
composition law involves $n$ elements keeping the order of nilpotency
the same. This generalization was studied by Mal'tsev and his collaborators
in 1960's.

Another possible generalization is to increase the order of nilpotency and
this generalization is our main concern in this paper. It is obvious that
this generalization requires algebras with at least ternary composition
law. Thus if ${\cal T}$ is an algebra with ternary multilplication
$(a,b,c)\to a\cdot b\cdot c\in {\cal T}$ then we shall call its
multiplication {\it ternary anti-commutative} if $a^3=0,\;\forall a\in
{\cal T}$. Then from the
identities $(a+b)^3=0$ and
$(a+b+c)^3=0$, where $a,b,c$
are an arbitrary elements of the algebra ${\cal T}$, it follows
immediately that
$$
a\cdot b\cdot c+b\cdot c\cdot a+c\cdot a\cdot b+
              c\cdot b\cdot a+a\cdot c\cdot b+b\cdot a\cdot c=0.
$$
\noindent
The left-hand side of the above identity suggests to introduce in analogy
with the classical anti-commutativity the ternary 
anti-commutator
\begin{equation}
\{a_1,a_2,a_3\}=\sum_{\sigma\in S_3}
         a_{\sigma(1)}\cdot a_{\sigma(2)}\cdot a_{\sigma(3)}. 
\label{anticomm}
\end{equation}
\noindent
If $a,b,c$ are the elements of some ternary algebra then we shall call
them ternary anti-commutative elements if $\{a,b,c\}=0$.

In this paper we study the algebras whose generators are ternary
anti-commutative. These algebras may be viewed as an analogues of
Grassmann algebra. Therefore we use the term ternary Grassmann algebra
for them.
Since classical Grassmann algebras have played an
essential role in supersymmetric field theories there have been made
attempts to find an applications of ternary Grassmann algebras in field
theories.  
The ternary Grassmann algebra with ternary defining relations is used in
[3] to construct the operators which are more fundamental then the
operators of supersymmetry.
The algebra with one ternary anti-commutative generator is used in [6] to 
construct the $Z_3$-graded quantum space and in [1], [7] to generalize the
algebras of supersymmetries. Therefore we hope that other ternary structures
such as ternary generalizations of Clifford and Lie algebras 
will be a ground for the field theories with new kind of symmetries.

%%%%%%%%%%%%%%%%%%%%%%%%%%%%%%%%%%%%%%%%%%%%%%%%%%%%%%%%%%%%%%%%%%%%%%%%%%
\vskip .6cm
\centerline{\tbf 2 Ternary Grassmann algebras}
\vskip .4cm
\indent
We begin this section with the general definition of the ternary Grassmann
algebra. An associative algebra over the field $\bf C$ generated by
$\theta_1,\theta_2,\ldots,\theta_N$ is called {\it ternary Grassmann
algebra} (TGA) if its generators satisfy the following condition of
ternary anti-commutativity:
\begin{equation}
\{\theta_A,\theta_B,\theta_C\}=0,\qquad \forall A,B,C=1,\ldots,N.
\label{TGA}
\end{equation}
\noindent
Since each classical Grassmann algebra is TGA we define {\it proper
ternary Grassmann algebra} (PTGA) as a ternary Grassmann algebra whose
generators satisfy the additional condition $\theta_A^2\not=0,\;\forall
A=1,\ldots,N$. 

The generators of any TGA are cubic nilpotent, i.e.
$\theta_A^3=0,\;\forall A=1,\ldots,N$. This property follows from
(\ref{TGA}) when $A=B=C$. PTGA can be endowed with the $Z_3$-grading
defined as follows: each generator $\theta_A$ has grade $1$ and the grade
of any monomial equals its degree with respect generators $\theta_A$
modulo $3$. 

We get the simplest example of a PTGA when $N=1$.
This algebra is a three dimensional vector space over the field $\bf C$
and it is spanned by the monomials $1,\theta,\theta^2$, where $\theta$ is
the generator. This algebra was used in [6] to construct the $Z_3$-graded
quantum space. 

In order to have an explicit construction of PTGA when $N>1$ one ought to
find the defining commutation relations which are consistent with
(\ref{TGA}). In this paper we shall describe two ways of solving the
condition (\ref{TGA}) of ternary anti-commutativity.

%%%%%%%%%%%%%%%%%%%%%%%%%%%%%%%%%%%%%%%%%%%%%%%%%%%%%%%%%%%%%%%%%%%%%%%%%%%%%%
\vskip .4cm
\centerline{\bf 2.1 Ternary Grassmann algebra with binary relations}
\vskip .3cm
\indent
Let us assume some 
binary commutation relations between the generators
$\theta_1,\ldots,\theta_N$ of PTGA. Let 
these binary relations be of the form
$$
\theta_A\theta_B=q_{AB}\,\theta_B\theta_A,
$$
\noindent
where $q_{AB}$ are complex numbers such that $q_{AB}\not=0$ for each pair
of indices $(A,B)$. It is clear that $q_{AB}=1$ for $A=B$ since
$\theta^2_A\not=0$ and $q_{AB}=q^{-1}_{BA}$. Putting these binary
commutation relations into the condition (\ref{TGA}) one obtains
\begin{equation}
1+q_{BA}+q_{CB}+q_{CA}\,q_{BA}+
             q_{CA}\,q_{CB}+q_{CB}\,q_{CA}\,q_{BA}=0.
\end{equation}
\noindent
If $B=C,\,B\not=A$ then the above condition takes on the form
$$
1+q_{BA}+q^2_{BA}=0,
$$
\noindent
which clearly shows that $q_{AB}$ is the cubic root of unit.
Here, we have a choice between $j$ and $j^2$, where $j=e^{{2\pi
i}\over 3}$. Let us choose $q_{AB}=j$ for $A>B$ and $q_{AB}=j^2$ for
$A<B$. It is obvious that another choice leads just to the same structure.
Now we are able to define the {\it PTGA with binary 
relations} between its generators.
This algebra is an associative
algebra over the field $\bf C$ generated by $\theta_1,\ldots,\theta_N$
which are subjected to the following commutation relations:
\begin{equation}
\theta_A\theta_B=q_{AB}\;\theta_B\theta_A,\qquad \theta^3_A=0,
\label{q-commutrel}
\end{equation}
where
\begin{equation}
q_{AB}=\left\{\begin{array}{ll}
  1,   & \mbox{$A=B$}\\
  j,   & \mbox{$A>B$}\\
  j^2, & \mbox{$A<B$}
 \end{array}
\right.
\label{definition-of-q}
\end{equation}
\indent
Let us denote this PTGA with binary relations by ${\cal G}^N_B$. 
In order to make the structure of the algebra ${\cal G}^N_B$ more
transparent we shall use generators with conjugate indices defined as 
$\theta_{\bar A}=\theta^2_A$. From commutation relations
(\ref{q-commutrel}) it follows then
\begin{equation}
\theta_A\theta_{\bar A}=\theta_{\bar A}\theta_A=0,\; 
 \theta_A\theta_{\bar B}=\bar q_{AB}\,\theta_{\bar B}\theta_A,\; 
   \theta_{\bar A}\theta_{\bar B}= q_{AB}\,\theta_{\bar B}\theta_{\bar A},\;
    \theta_{\bar A}^2=0.
\end{equation}
It is helpful to introduce notations which are similar to Kostant ones for
the clas\-si\-cal Grass\-mann al\-geb\-ra.
Let ${\cal N}=\{1,\ldots,N\}$ and $J=(A_1,\ldots,A_k)$
be a subset of ${\cal N}$. We associate two monomials $\theta_J$ and
$\theta_{\bar J}$ to each subset $J\subset {\cal N}$ defining them as
follows: 
\begin{equation}
\theta_J=\theta_{A_1}\theta_{A_2}\ldots\theta_{A_k},\quad
   \theta_{\bar J}=\theta_{{\bar A}_1}\theta_{{\bar A}_2}
          \ldots\theta_{{\bar A}_k}.
\end{equation}
If $J=\emptyset$ then as usual $\theta_{\emptyset}=1$. Then the algebra
${\cal G}^N_B$ is a vector space over ${\bf C}$ spanned by the monomials
$\theta_J\,\theta_{\bar K}$ such that $J\cap K=\emptyset$. Thus an
arbitrary element $f(\theta)$ of ${\cal G}^N_B$ can be expressed as
\begin{equation}
f(\theta)=\sum_{J\cap K=\emptyset} \alpha_{J\bar K}\,
                               \theta_{J}\,\theta_{\bar K},
\end{equation}
where $\alpha_{J\bar K}$ are complex numbers.

The number of subsets $J\subset {\cal N}$ consisting of $k$ elements is
$C^k_N$. Since the subset $K\subset {\cal N}$ matches $J$ if $J\cap
K=\emptyset$ it is obvious that $K\subset {\cal N}\backslash J$ and the
number of such subsets is $2^{N-k}$. Thus the total dimension of ${\cal
G}^N_B$ is $\sum^{N}_{k=1} C^k_N\,2^{N-k}=3^N$.
The highest degree monomial of the algebra
${\cal G}^N_B$ is the monomial 
$\theta_{\bar 1}\theta_{\bar 2}\ldots\theta_{\bar
                              N}=\theta^2_1\theta^2_2\ldots\theta^2_N$. 

If we write an arbitrary element $f(\theta)$ in the form
$f(\theta)=\alpha+O(\theta)$,
where $O(\theta)$ stands for the
terms each containing at least one generator $\theta_A$ 
then it can be shown that there exists the inverse element
$f^{-1}(\theta)$ if and only if $\alpha\not=0$. 
%%%%%%%%%%%%%%%%%%%%%%%%%%%%%%%%%%%%%%%%%%%%%%%%%%%%%%%%%%%%%%%%%%%%%%%%%%%%
\vskip .4cm
\centerline{\bf 2.2. Ternary Grassmann algebra with ternary relations}
\vskip .3cm
\indent
Another way to solve the conditions of ternary anti-commutativity
(\ref{TGA}) is to assume some ternary commutation relations between
generators $\theta_A$. Since the cyclic subgroup $Z_3$ of the group $S_3$
has the representation by cubic roots of unit it seems natural in analogy
with the ordinary anti-commutativity to
construct ternary commutation relations by means of the action of the
cyclic group $Z_3$ on the indices of the corresponding variables.
This idea was first proposed by R. Kerner and in this subsection we
briefly describe the structure of the corresponding ternary Grassmann
algebra. More detailed description of the ternary Grassmann algebra with
ternary relations and its applications can be found in [2], [3].

The {\it PTGA with ternary commutation relations} is an associative
algebra over the field ${\bf C}$ generated by
$\theta_1,\theta_2,\ldots,\theta_N$ which 
are subjected to the following ternary defining relations:
\begin{equation}
 \theta_{A} \theta_{B} \theta_{C} = j\, \theta_{B} \theta_{C} \theta_{A}.
 \label{ternrel}
\end{equation}
\noindent
Let us denote the PTGA with ternary relations by ${\cal G}^N_T$. The above
ternary defining relations (\ref{ternrel}) are based on the idea of the
action of the cyclic group $Z_3$ in the sense that each cyclic permutation
of the indices in the product $\theta_A\,\theta_B\,\theta_C$ is
accompanied by the multiplication by cubic root of unit accordingly to the
representation of $Z_3$. It is obvious that the generators of ${\cal
G}^N_T$ satisfy the conditions (\ref{TGA}) of ternary anti-commutativity.

It should be noted here that there are no any relations between the binary
products $\theta_{A} \theta_{B}$ of generators of ${\cal G}^N_T$ that is
they are linearly independent entities.
The immediate corollary from the above definition is that any product of
four or more generators must vanish. Here is the proof: 
\begin{eqnarray*}
\lefteqn{(\theta_{A} \theta_{B} \theta_{C}) \theta_{D}
                               = j \theta_{B}(\theta_{C} 
                                     \theta_{A} \theta_{D})
= j^2 (\theta_{B} \theta_{A} \theta_{D})\theta_{C}}\qquad\qquad
\ \ \ \ \ \ \
\ \ \ \ \ \ \ \ \\ & & 
\ \ \ \ \ = \theta_{A}(\theta_{D} \theta_{B} \theta_{C}) = j \theta_{A}
\theta_{B} \theta_{C} \theta_{D}.
\end{eqnarray*}
Now, as $(1 - j) \not = 0$, one must have $\theta_{A} \theta_{B} \theta_{C}
\theta_{D} = 0$. Thus the monomials $\theta_A\,\theta^2_B$ are the
highest degree monomials of the algebra ${\cal G}^N_T$.
The dimension of the PTGA with ternary relations is ${N(N+1)(N+2)/3}+1 $,
because we have $N$ generators, $N^2$ independent products of two generators,
$N(N-1)$ independent ternary expressions with two generators equal and one
different, and $N(N-1)(N-2)/3 $ ternary products with all the three
generators different; finally, the numbers give an extra dimension.
Any cube of a generator is equal to zero; the odd permutation of factors
in a product of three leads to an independent quantity. 
%%%%%%%%%%%%%%%%%%%%%%%%%%%%%%%%%%%%%%%%%%%%%%%%%%%%%%%%%%%%%%%%%%%%%%%%%%%%%
\vskip .6cm
\centerline{\tbf 3 Integration}
\vskip .4cm
\indent
The aim of this section is to define the derivatives and integral over the
PTGA generated by an arbitrary number of
generators $N$. We shall also establish and prove the formula of a change
of variables in the integral over the ternary Grassmann algebra. Though
the definitions of derivatives and integrals are just the same in  
both cases of ${\cal G}^N_B$ and ${\cal G}^N_T$ we shall always assume in
this section that we are considering the algebra ${\cal G}^N_B$.
%%%%%%%%%%%%%%%%%%%%%%%%%%%%%%%%%%%%%%%%%%%%%%%%%%%%%%%%%%%%%%%%%%%%%%%%%%%%%
\vskip .4cm 
\centerline{\bf 3.1 Derivatives and integral}
\vskip .3cm
\indent
Using the notations of the subsection 2.1 we define the derivatives with
respect generators $\theta_A$ by the following set of rules: 
\begin{equation}
\partial_{A}(\theta_B)=\delta_{AB},\quad 
    \partial_{A}(\theta_{\bar B})=(1+j^2)\,\delta_{AB}\theta_B.
\label{derivatives}
\end{equation}
\noindent
It is also helpful to define the derivatives with respect squares of
generators $\theta_{\bar A}=\theta^2_A$ as follows:
\begin{equation}
\partial_{\bar A}(\theta_B)=0,\quad 
   \partial_{\bar A}(\theta_{\bar B})=\delta_{AB}.
\label{square-derivatives}
\end{equation}
\noindent
It is easy to establish the relation between these derivatives and the
second order derivatives
$$
\partial_{\bar A}=(1+j)\;\partial^2_{A}.
$$
\noindent
Clearly that each derivative $\partial_{A}$ is an operator of  
grade $2$ and each derivative $\partial_{\bar A}$ is an operator of 
grade $1$. 
Straightforward computation shows that derivatives satisfy the following
commutation relations:
$$
\partial_A^3=0,\ \ \ 
\partial_A\,\partial_{\bar A}=\partial_{\bar A}\,\partial_{A}=0,\ \ \ \ 
          \partial_A\,\partial_B=q_{AB}\,\partial_B\,\partial_A,
$$
$$
\partial_A\,\partial_{\bar B}=
          {\bar q}_{AB}\,\partial_{\bar B}\,\partial_A,\ \ \ 
\partial_{\bar A}\,\partial_{\bar B}=
           q_{AB}\,\partial_{\bar B}\,\partial_{\bar A},\ \ \ \ 
\partial_{\bar A}^2=0.
$$
\noindent
From the above formulae it follows that derivatives
$\partial_A,\,\partial_{\bar A}$ are ternary
anti-commutative, 
i.e.
$$
\{\partial_{A},\partial_{B},\partial_{C}\}=0,\quad
   \{\partial_{\bar A},\partial_{\bar B},\partial_{\bar C}\}=0.
$$    

Integral over the PTGA generated by one generator 
was defined and studied in [6]. We extend there given definition to the
PTGA with $N$ generators and prove the formula of a change of
variables. The integral of an arbitrary element $f(\theta)\in {\cal
G}^N_B$ with respect $\theta_A$ is defined by the formula 
\begin{equation}
\int d\theta_A\; f(\theta)=
                 \partial_{\bar A}(f(\theta)).
\label{integral}                 
\end{equation}
\noindent     
As usual the multiple integral is to be understood as the repeated integral. 

Note that integration with respect all generators in the case of the PTGA
with ternary relations ${\cal G}^N_T$ is always trivial since the highest
degree monomials have the form $\theta_A\theta^2_B$.
Integration with respect all generators in the case of PTGA with binary
relations  ${\cal G}^N_B$ yields the coefficient 
at the highest degree monomial. Thus
\begin{equation}
\int {\cal D}\theta\; f(\theta)=
                 \alpha_{\bar 1\bar 2\ldots \bar N},
\label{N-integral}
\end{equation}
\noindent
where ${\cal D}\theta=d\theta_1\,d\theta_2\ldots d\theta_N$ and 
$\alpha_{\bar 1\bar 2\ldots \bar N}$ is the coefficient at the monomial
$\theta_{\bar 1}\theta_{\bar 2}\ldots \theta_{\bar N}$.

Let $\vartheta_1,\vartheta_2,\ldots,\vartheta_N$ be another system
of generators of the algebra ${\cal G}^N_B$ and 
generators $\theta_1,\theta_2,\ldots,\theta_N$ are expressed in terms
of $\vartheta_1,\vartheta_2,\ldots,\vartheta_N$ as follows:
\begin{equation}
\theta_A=\sum_{B=1}^N \alpha_{AB}\,\vartheta_B+O_{A}(\theta),                              
\label{transformation}
\end{equation}
\noindent
where $O_A(\theta)$ means terms containing more than one generator and the
determinant of the matrix $A=(\alpha_{AB})$ differs from zero.
If $T(\theta,\vartheta)$ is the Jacobian matrix of
the above transformation then
we define the {\it Jacobian} $J(\theta,\vartheta)$ 
by the formula
\begin{equation}
J(\theta,\vartheta)=det^{-2}(T(\theta,\vartheta)).
\label{jacobian}   
\end{equation}
\noindent
It can be proved that 
\begin{equation}
\int {\cal D}\theta\; f(\theta)=
   \int {\cal D}\vartheta\; J(\theta,\vartheta) \tilde f(\vartheta).
\label{change-of-variables}
\end{equation}
It should be noted that in contrast to fermion integral where determinant
of the Jacobian matrix appears in the formula of a change of variables in
power $-1$ in the above formula it has the power $-2$.

Now we turn to the proof of the formula (\ref{change-of-variables}). It is
based on the observation that if two systems of generators of the algebra
${\cal G}^N_B$ are related by the formulae (\ref{transformation}) then
this imposes (in the contrast to the classical 
Grassmann algebra) very strong restrictions on the coefficients of the
expressions at the right-hand sides of (\ref{transformation}). Since these
restrictions lead to the bulky conditions on the coefficients we produce 
only the conditions for the entries of the matrix $A$ and prove the
formula (\ref{change-of-variables}) when the right-hand side expressions of
(\ref{transformation})  contain only linear terms with respect generators.
The entries of $A$ must satisfy the following conditions:
\begin{equation}\left
  \{\begin{array}{ll}
\alpha_{AD}\alpha_{BC}=0 & \mbox{$A<B,\, C<D$}\\
\alpha_{AD}\alpha_{BD}=0 & \mbox{$A<B$, (no summation!).}
  \label{conditions}
   \end{array}\right.
\label{conditions-for-coefficients}
\end{equation}
\noindent
Taking into account also the condition $det\,A\not=0$
guaranteering the linear independce of the new generators we conclude that
the matrix $A$ is a diagonal matrix. 
Thus we have
$$
\int {\cal D}\vartheta\, J(\theta,\vartheta)\,\tilde f(\vartheta)=
  \int {\cal D}\vartheta\, \prod_{A=1}^N 
               (\alpha^{-2}_{AA})\;\tilde f(\vartheta)=
                    \int {\cal D}\theta\, f(\theta),
$$
\noindent
and this ends the proof.
%%%%%%%%%%%%%%%%%%%%%%%%%%%%%%%%%%%%%%%%%%%%%%%%%%%%%%%%%%%%%%%%%%%%%%%%%%%
\vskip .4cm
\centerline{\bf 3.2 Pfaffian of a cubic matrix}
\vskip .3cm
\indent
It is well-known ([5]) that the fermion integral of Gaussian type over the
even dimensional 
classical Grassmann algebra can be used to derive the Pfaffian of a
skew-symmetric square matrix. Replacing the notion of a
skew-symmetric square matrix by its cubic analogue and making use of the
integral over PTGA with binary relations 
we define the Pfaffian of a cubic matrix and calculate its explicit form 
in the dimension $N=3$. 

Let $\Omega(\theta)$ be a cubic form 
\begin{equation}
\Omega(\theta)={1\over 3}\,\omega_{ABC}\,\theta_A\theta_B\theta_C,
\label{cubic-form}
\end{equation}
\noindent
with the coefficients satisfying the relations
$$
\omega_{ABC}={\bar q}_{AB}\,\omega_{BAC},\;
      \omega_{ABC}={\bar q}_{BC}\,\omega_{ACB},\;
$$
\noindent
if there are at least two different indices in the triple $(A,B,C)$ and
$$
\omega_{AAA}=0.
$$
\noindent
The coefficients $\omega_{ABC}$ of the
cubic form $\Omega(\theta)$ can be considered as the entries of a cubic
$N\times N\times N$-matrix we shall denote by $\Omega$. 
From the above relations it follows that the entries of the cubic matrix
$\Omega$ satisfy the relations
\begin{equation}
\omega_{ABC}+\omega_{BCA}+\omega_{CAB}+\omega_{BAC}+
    \omega_{ACB}+\omega_{CBA}=0,
\label{cubic-skew-symmetry}
\end{equation}
\noindent
for any triple of indices $A,B,C$. The property
(\ref{cubic-skew-symmetry}) may be considered as a cubic generalization of
the notion of skew-symmetric square matrix. 
We define the {\it
Pfaffian} of this cubic matrix by the following integral:
\begin{equation}
Pf_{cub}(\Omega)=\int {\cal D}\theta\; e^{\Omega(\theta)}.
\label{definition-of-Pfaffian}
\end{equation}
\noindent
It is not a surprise that the above integral leads to a non-trivial result
only when the number $N$ of generators 
is a number divisible by $3$. Thus the dimension $N=3$ is the
lowermost dimension providing a non-trivial result. Let us find the
Pfaffian of a cubic matrix in this case. The cubic form (\ref{cubic-form})
then takes on the form
\begin{eqnarray*}
\Omega(\theta)&=& 2\,\omega_{123}\,\theta_1\theta_2\theta_3+
  \omega_{112}\,\theta^2_1\theta_2+\omega_{122}\,\theta_1\theta^2_2+
   \omega_{113}\,\theta^2_1\theta_3\\
& &\mbox{\ \ \ \ \ \ \ \ \ \ \ \ \ \ \ \ \ \
\ }+\omega_{133}\,\theta_1\theta^2_3+ 
    \omega_{223}\,\theta^2_2\theta_3+\omega_{233}\,\theta_2\theta^2_3.
\end{eqnarray*}
\noindent
Making use of the
definition of the integral over ternary Grassmann 
algebra one obtaines the following homogeneous polynomial for the
Pfaffian: 
\begin{eqnarray*}
Pf_{cub} (\Omega)&=& \int {\cal D}\theta\; e^{\Omega(\theta)}=
   \int {\cal D}\theta\; (1+\Omega(\theta)+{1\over 2!}\Omega^2(\theta))\\
  & & \mbox{}=
    4\,\omega^2_{123}-\omega_{211}\omega_{233}-\omega_{221}\omega_{133}
       -\omega_{311}\omega_{223}.
\end{eqnarray*}
\noindent
We end this section by the following speculation. If we would develop the
calculus of the cubic matrices based on the ternary Grassmann algebra 
approach we would define the determinant of
the cubic $3\times 3\times 3$-matrix $\Omega$ as the third power of the
above polynomial $Pf_{cub}(\Omega)$ and then the determinant would be a sum of
products each containing six entries of the cubic matrix $\Omega$. This
suggests that the group $S_3$ of permutations of three elements is likely
to play an essential role in the definition of the detrminant of cubic
matrices.
%%%%%%%%%%%%%%%%%%%%%%%%%%%%%%%%%%%%%%%%%%%%%%%%%%%%%%%%%%%%%%%%%%%%%%%%%%%%
\vskip .6cm
\centerline{\tbf Acknowledgements}
\vskip .4cm
\indent
I am grateful to Prof. R. Kerner for explaining to me his
ideas on $Z_3$-graded and ternary structures. I express my thanks to
Dr. R. Roomeldi for pointing out to me the algebraic literature
concerning the notion of anti-commutativity and for computer programs
helping to analyze the structure of ternary Grassmann algebras.
This work was partly supported by the Estonian Science Foundation 
(grant nr. 368).                               
%%%%%%%%%%%%%%%%%%%%%%%%%%%%%%%%%%%%%%%%%%%%%%%%%%%%%%%%%%%%%%%%%%%%%%%%
\newpage
\vskip .4cm
\centerline{\tbf References}

{\parindent=0pt
$$\vbox {\halign{#\hfil \quad & \vtop {\hsize=31em\strut #\strut }\cr
1.& V. Abramov, {\it $Z_3$-graded analogues of Clifford al\-geb\-ras and
al\-geb\-ra of $Z_3$-gra\-ded symm\-et\-ries}, Algebras, Groups \&
Geometries, Vol. 12, No. 3, p. 201-221, 1995.\cr
2.& V. Abramov, R. Kerner, B. Le Roy, {\it Hypersymmetry: a $Z_3$-graded
generalization of supersymmetry}, to appear in J. Math. Phys.\cr
3.& R. Kerner, J. Math. Phys., 1992, {\bf 33}(1), 403.\cr
4.& B. Le Roy, C.R.Acad. Sci. Paris, 1995, {\bf 320}, 593-598. \cr
5.& V. Mathai, D. Quillen, Topology, 1986, {\bf 25}(1), 85-110.\cr 
6.& Won-Sang Chung, J. Math. Phys., 1994, {\bf 35}(5), 2497-2504.\cr
7.& Noureddine Mohammedi, preprint SHEP 94/95-17, December, 1994.\cr
8.& Jaime Keller, In: Spinors, Twistors, Clifford Algebras and Quantum
Deformations, 1993, 189-196.\cr}}$$}

\end{document}